\documentclass[useAMS,usenatbib]{mn2e}
\usepackage{graphicx}

\def\ltorder{\mathrel{\raise.3ex\hbox{$<$}\mkern-14mu\lower0.6ex\hbox
{$\sim$}}}
\def\gtorder{\mathrel{\raise.3ex\hbox{$>$}\mkern-14mu\lower0.6ex\hbox
{$\sim$}}}

\title[HI Detection of Clouds and Galaxies]
  {The HI Detection of Low Column Density Clouds and Galaxies}
\author[Suzanne M.~Linder et al.]
  {Suzanne M.~Linder$^1$, Robert F.~Minchin$^1$, Jonathan I.~Davies$^1$, 
Maarten Baes$^{1,2}$
\newauthor
Rhodri Evans$^1$, Sarah Roberts$^1$, Sabina Sabatini$^1$, and Rodney Smith$^1$
\\$^1$Cardiff University, Queen's Buildings, 5, The
Parade, Cardiff CF24 3YB, United Kingdom\\
$^2$Sterrenkundig Observatorium, Universiteit Gent, Krijgslaan 281-S9, B-9000 
Gent, Belgium
      }
\date{Received 2003 Xxxxx XX}

\pagerange{\pageref{firstpage}--\pageref{lastpage}} \pubyear{2003}

\def\LaTeX{L\kern-.36em\raise.3ex\hbox{a}\kern-.15em
    T\kern-.1667em\lower.7ex\hbox{E}\kern-.125emX}

\begin{document}

\label{firstpage}

\maketitle

\begin{abstract}
The HIDEEP survey (Minchin et al.~2003) was done in an attempt to find objects having low
inferred neutral hydrogen column densities, yet they found a distribution which was
strongly peaked at $10^{20.65}$ cm$^{-2}$.  In an attempt to understand this distribution
and similar survey results, we model HI profiles of gas discs and use simple simulations
of objects having a wide range of HI properties in the presence of an ionizing background.
We find that inferred
column density ($N_{HI}^o$) values, which are found by averaging total HI masses over some disc
area,
do not vary strongly with central column 
density ($N_{max}$) for detectable objects,
so that even a population having a wide range of $N_{max}$ values
will give rise to a strongly peaked distribution of $N_{HI}^o$ values.
We find that populations of objects, having a wide range of model parameters,
give rise to inferred column density distributions around $10^{20.6\pm 0.3}$ cm$^{-2}$.
However, populations of fairly massive objects having a wide range of central column densities
work best in reproducing the HIDEEP data, and these populations are also consistent with 
observed Lyman limit absorber counts.  It may be necessary to look two orders of magnitude
fainter than HIDEEP limits to detect ionized objects having central column densities $<10^
{20}$ cm$^{-2}$, but the inferred column densities of already 
detected objects might
be lower if their radii could be estimated more accurately.
\end{abstract}


\section{Introduction}
Understanding the properties of dwarf galaxies, large diffuse galaxies, and
any clouds of 
similar mass is important in understanding the formation of galaxies.
For example, Cold Dark Matter theory predicts the existence of a population 
of low-mass satellite galaxies
(e.g. Moore et al. 1999; Klypin et al. 1999).
Furthermore, studying the properties of such objects
is important in understanding the
nature of Ly$\alpha$ absorbers and metal line absorbers such as weak MgII
systems (Rigby, Charlton, \& Churchill 2002).
Any such objects which have yet been undetected may have a different range of averaged
neutral hydrogen column densities from that of the known population of galaxies.

Gas having a wide range of neutral column density ($N_{HI}$) values has
been observed as Ly$\alpha$ absorption at low redshifts (for example, Bahcall et al.~1996)
where absorption lines shortward of Ly$\alpha$ emission in quasar
spectra arise from lines of sight through intervening gas between us and
the quasar, and $N_{HI}$ ranges from $<10^{12}$ cm$^{-2}$ to $\sim 10^{21}$
cm$^{-2}$.  Larger amounts of gas with $N_{HI}\gtorder 10^{19}$ cm$^{-2}$
can also be observed more directly as 21 cm emission in the local universe.
The strongest `damped' Ly$\alpha$ absorbers, with  $N_{HI} > 10^{20.3}$
cm$^{-2}$, are often found to arise in lines
of sight through galaxies including several low surface brightness (LSB)
and dwarf galaxies
(Cohen 2001; Turnshek et al.~2000; Bowen, Tripp, \& Jenkins 2001).
Yet the somewhat weaker Lyman limit systems ($N_{HI} > 10^{17.2}$ cm$^{-2}$),
the column densities of which are more difficult to measure accurately,
have long been thought to arise in lines of sight through luminous galaxies
(Bergeron \& Boiss\'e 1991; Steidel 1995).
Some weaker Ly$\alpha$ forest absorbers are thought to arise in small
amounts of intergalactic gas (Dav\'e et al.~1999), while some could arise in
gas surrounding galaxies (e.g. Chen et al.~2001; Linder 1998; 2000).

A recent HI survey (HIDEEP; Minchin 2001; Minchin et al. 2003) was capable
of detecting objects with inferred neutral hydrogen column densities
($N_{HI}^o$) as low as $4\times 10^{18}$ cm$^{-2}$ for galaxies
having velocity width $\triangle V=200$ km s$^{-1}$, assuming that a galaxy
with suitable properties fills the telescope beam.  Yet they failed to find 
anything
with $N_{HI}^o<10^{20}$ cm$^{-2}$.  Other HI surveys have also
found that galaxies show little variation in column densities averaged
over some radius (Zwaan et al.~1997), although the integration times may 
not be long enough to detect low column density galaxies in such surveys, as
discussed by Minchin et al.~(2003).  These HI surveys are limited by flux,
rather than column density, when detecting faint objects, and the column 
density of the detected objects is uncertain given that the sources are
generally unresolved.  However there is a limit on column density in a survey
such as HIDEEP in the sense that a resolved, low column density object
could fill the beam, although such objects are not often seen.

Rosenberg \& Schneider (2003) found that their sample of
HI-selected galaxies obey a relationship between HI cross section 
and HI mass, which is equivalent to having fairly
constant averaged column densities.
They plot, in their first figure,
the disc areas $A_{DLA}$, where $N_{HI}>2 \times 10^{20}$  cm$^{-2}$ and 
thus where
damped Ly$\alpha$ absorbers can arise, versus the HI mass ($M_{HI}$) for a
sample of  HI selected galaxies.
Some scatter is seen in the log-log plot, yet they can easily fit a line having a slope of
about one. Thus they find $\log(A_{DLA})=\log(M_{HI})-6.82$,
which would imply that galaxies 
having a wide range of mass and HI sizes all have area-averaged column densities 
of around $8\times 10^{20}$  cm$^{-2}$, where the displayed points are all within
about $0.8$ orders of magnitude from the fitted line. 

Similar correlations between HI size and HI mass have also been
seen by Giovanelli \& Haynes (1983) and Verheijen \& Sancisi (2001), and
a correlation between HI mass and optical sizes of galaxies has also been seen
by Haynes \& Giovanelli (1984).
Other surveys, capable of
detecting low HI mass objects at various sensitivities, including some
directed toward detecting
extragalactic
high velocity clouds (HVCs) (Blitz et al.~1999; Charlton, Churchill \& Rigby 2000;
Davies et al.~2002) have been largely
unsuccessful at finding objects with low HI masses
(de Blok et al.~2002; Zwaan \& Briggs
2000; Dahlem et al. 2001; Zwaan 2001; Verheijen et al.~2000).  On the other
hand, some very faint optical sources have been found to be rich in gas
(Davies et al.~2001), and there is theoretically no reason to expect every HI cloud
to be capable of forming large amounts of stars.
Furthermore, small HVCs with peak $N_{HI}\sim 6\times 10^{18}$ cm$^{-2}$
are being detected around our Galaxy (Hoffman, Salpeter, \& Pocceschi 2002)
and around M31 (Thilker et al.~2004).

One suggested explanation for the lack of low column density detections
in the HIDEEP survey is that the gas is hidden in 'frozen discs' (Minchin et al.~2003)
where the 21 cm transition is not excited to a spin temperature
above the cosmic background (Watson \& Deguchi 1984).

A second possible explanation for the lack of low column density
detections is that the gaseous discs become highly ionized at a disc radius
not far beyond that where $N_{HI}=10^{20}$  cm$^{-2}$, so that the average
inferred value remains above $10^{20}$ cm$^{-2}$.  The ionization of outer
galaxy discs by a background of Lyman continuum photons
was suggested and modelled first by
Bochkarev \& Sunyaev (1977) and later by Maloney (1993), Dove \& Shull
(1994a), and Corbelli \& Salpeter (1993) in order to explain the sudden
truncations seen in carefully observed spiral galaxy discs.
Since then ionized gas has been detected in H$\alpha$ emission using a
Fabry-Perot `staring technique' (Bland-Hawthorn et al.~1994) beyond the
HI edges of several nearby galaxies
(Bland-Hawthorn, Freeman \& Quinn 1997; Bland-Hawthorn 1998).  The ionizing
background has been measured most recently at low redshifts by Scott et 
al. (2002), who find $J(912$\AA$)=7.6^{+9.4}_{-3.0}\times10^{
-23}$
erg cm$^{-2}$ s$^{-1}$ Hz$^{-1}$ sr$^{-1}$.
Gas in the ionized parts of outer galaxy discs is likely to give rise to
at least some Ly$\alpha$ absorption (Linder 1998; 2000), and some variations
will arise in the column density value at which HI discs are truncated as
a result of fluctuations in the ionizing background radiation (Linder et 
al.~2003).

Ionized gas clouds cannot correctly be referred to as undetected 'HI clouds'
(although current HI structures may have been ionized in the recent cosmological
past).  However structures containing ionized gas are interesting and relevant
to the galaxy formation process.  For example, ionized gas contains enough 
neutral atoms to give rise to all of the Lyman alpha absorbers (except for the
damped ones), and is thus, in principle, detectable in deep HI observations.
HVCs may also contain mostly ionized gas.  It is unknown whether massive clouds
exist far from luminous galaxies, although not all of the absorbers arise close
to galaxies (Stocke et al.~1995).  Ionized gas clouds may have small regions containing
HI clouds if the gas is sufficiently clumpy.

In this paper, we wish to understand
the observed lower limits in averaged column densities, and to 
constrain the properties of
any objects that could be going undetected in HI surveys as a result of
photoionization.  Section 2 discusses the modelling of HI discs and calculation
of column densities from HI observations.  Section 3 describes the method used 
to model galaxy and cloud
HI profiles and simulate populations of objects having a wide range of properties
in the presence of an ionizing background.
The results of such simulations are discussed in Section 4.
The value for the Hubble constant is assumed to be
$H_0=80$ km s$^{-1}$ Mpc$^{-1}$.

\section{HI Discs and Averaged Column Densities}

\begin{figure}
\vspace{0.5cm}
\includegraphics[width=84mm]{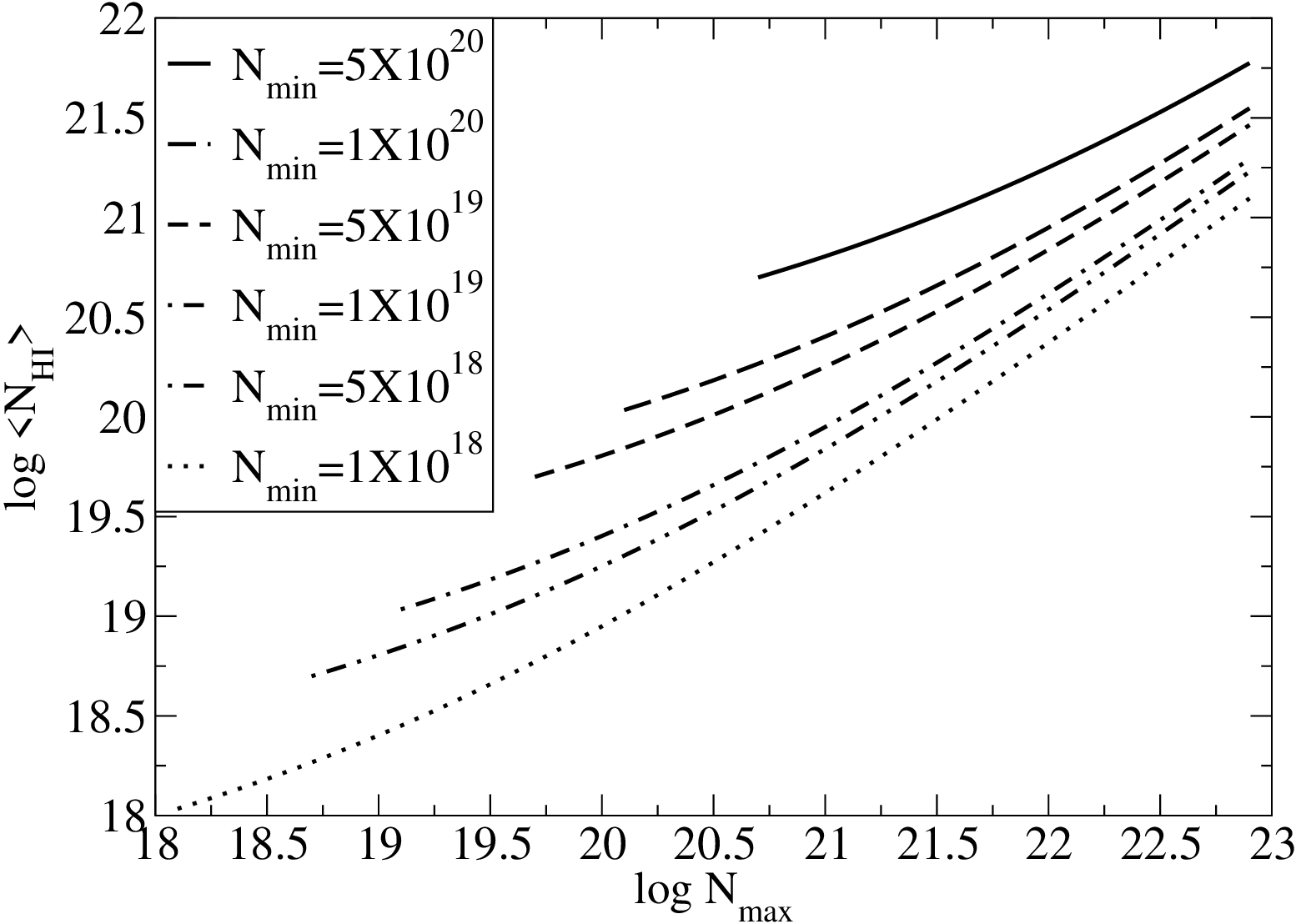}
 \caption{Values of the average column density 
$\langle N_{HI}\rangle$ are plotted versus central column density $N_{max}$, as in 
equation (2),
for several values of limiting $N_{min}$.  The curves are fairly flat for lower values of
$N_{max}$.  Thus fairly constant observed column densities would allow for
$N_{max}$ to have a wide range of moderate to low values.  The formula (2) plotted
here is valid only for purely exponential discs, without considering ionization effects.
}
\label{nhavgnmaxmin}
\end{figure}

Suppose all galaxies have exponential HI column density profiles (as modeled 
for example, by Swaters et al.~2002)
with central column density $N_{max}$ and scale length $h$ so that $N_{HI}(r)=
N_{max}exp(-r/h)$ at a radius $r$ along the disc.  If an observer 
maps the profile out to column density $N_{min}$, or an equivalent radius of
$h\ln (N_{max}/N_{min})$, the HI mass could then be 
found within an area $A=\pi h^2 \ln^2(N_{max}/N_{min})$.  The HI mass within this
radius, where $N_{HI}>N_{min}$, would be 
\begin{equation}
M_{HI}=2\pi h^2m_HN_{max}\left[1-\frac{N_{min}}{N_{max}}\left(1+\ln\frac{N_{max}}{N_{min}}
\right)\right]
\end{equation}
where $m_H$ is the mass of a hydrogen atom.
The averaged column density $\langle N_{HI}\rangle= M_{HI}/(m_HA)$ thus becomes
\begin{equation}
\langle N_{HI}\rangle = 2N_{max}\frac{\left[1-\frac{N_{min}}{N_{max}}\left(1+\ln\frac{N_{max}}{N_{min}}
\right)\right]}{\ln^2\left(\frac{N_{max}}{N_{min}}\right)}
\end{equation}
which depends only on
$N_{max}$ for a sample of galaxies mapped out to a constant $N_{min}$, and not on the 
scale length $h$.  However, a fairly constant 
$\langle N_{HI}\rangle$ does not imply a constant $N_{max}$, as seen in Fig.~\ref{nhavgnmaxmin}, where
$\langle N_{HI}\rangle$ is plotted versus $N_{max}$, as in equation (2),
for several values of $N_{min}$.  A narrow range of $\langle N_{HI}\rangle$ allows for
a somewhat wider range of $N_{max}$ and only implies that $N_{max}$ values are not likely to be
in the higher range shown on the plot, where the curves become steeper.  (We also know
that damped Ly$\alpha$ absorbers are not seen with $N_{HI}\gtorder 10^{22}$ cm$^{-2}$.)

Some scatter
is seen in the column density measurements from Minchin et al.~(2003) and Rosenberg \& Schneider (2003), and
in central column densities found in studies of LSB galaxies such as de Blok, McGaugh, 
\& van der Hulst (1996).  Thus some variation is likely to occur for values of 
$N_{max}$, though its
distribution is unknown.

Note that while the Rosenberg \& Schneider (2003) data can be compared with
the $\langle N_{HI}\rangle$ values shown 
above, Minchin et al.~(2003) calculate 
column densities somewhat differently, so we will refer to these as `inferred'
column densities ($N_{HI}^o$) in order to distinguish from the averaged ones 
discussed above.  
In this case the complete HI mass of a galaxy is 
measured, assuming that the mass is entirely within the beam.  Since they 
have not resolved their detected objects or obtained
column density profiles, they use an estimate of the HI 
radius in order to find an inferred column density.  They assume that the
HI radii are equivalent to 
5 times their observed effective optical radii, based upon the relationships from 
Salpeter \& Hoffman (1996).  Here we generally assume that such a radius corresponds to 
that where $N_{HI}=10^{20}$ cm$^{-2}$, although Salpeter \& Hoffman (1996) 
base their relationships on several studies which may have slightly varying limiting
values.  Since the $N_{HI}^o$ values are found differently from $\langle N_{HI}\rangle$ 
values, they
will depend here on what fraction of the HI mass
is contained within the estimated radius.  The inferred column density for an exponential
disc where the complete mass is accurately measured would be
\begin{equation}
N_{HI}^o = \frac{N_{max}}{\ln^2(N_{max}/N_{min})}.
\end{equation}
In this case $N_{min}$ is simply the column density corresponding to the estimated
galaxy radius.
Like equation (2), this formula does not depend upon the disc scale length.  

\begin{figure}
\vspace{0.5cm}
\includegraphics[width=84mm]{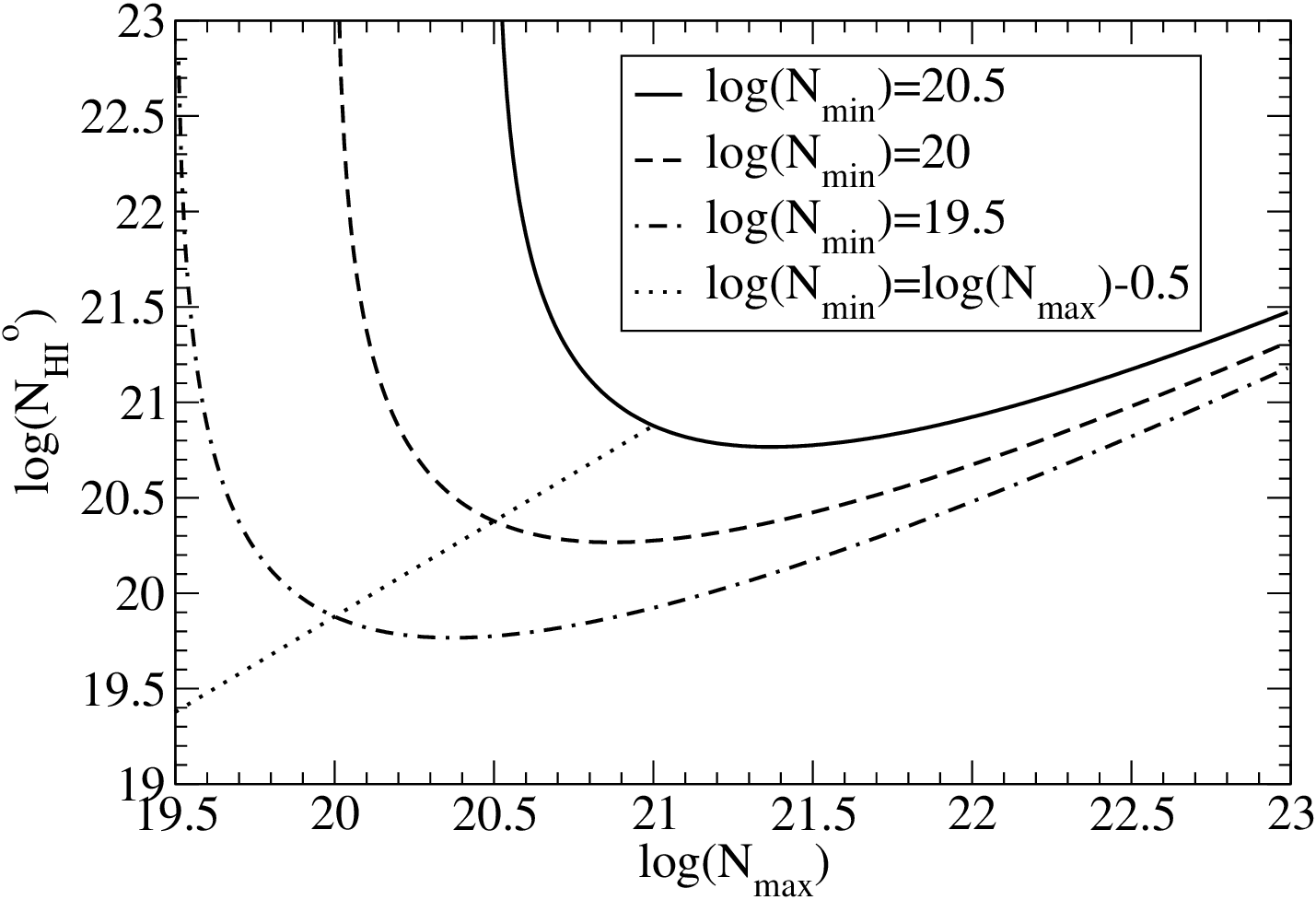}
 \caption{Values of inferred column density $N_{HI}^o$ are plotted versus central
column density $N_{max}$, as in 
Equation (3), again for exponential discs without yet considering ionization,
for several values of limiting $N_{min}$.  Here the curves turn up for lower values of
$N_{max}$ when a constant $N_{min}$ is assumed, as most of the mass will be 
outside the assumed radii for these objects.  Thus there is a minimum 
$N_{HI}^o$ that can be measured for a given $N_{min}$, and lower $N_{HI}^o$
values can only be found if the radii are estimated differently for objects
having low $N_{max}$ values, for example as shown in the line where $\log
N_{min} = \log N_{max}-0.5$.
}
\label{nhinfvsnmax}
\end{figure}

Objects having $N_{max}\sim N_{min}$ will have more of their mass outside
the radius at $N_{min}$, which will cause $N_{HI}^o$ to become large
as seen in Fig.~\ref{nhinfvsnmax}.  Thus there is a minimum $N_{HI}^o$ value
that can be detected for a given constant $N_{min}$.  Objects having 
$N_{max}\sim N_{min}$ are likely to have small radii however, and these radii
may be difficult to measure in a manner consistent with that used for larger
objects.  Thus an observer might effectively be using a smaller $N_{min}$ for a smaller 
$N_{max}$ when estimating any measurable radius for such an object. For example if
$\log N_{min}=\log N_{max}-0.5$ then $N_{HI}^o=0.745 N_{max}$ for an exponential disc, as 
shown by the dotted line in Fig.~\ref{nhinfvsnmax}.  

Different galaxy radii are used in different studies, inside of which the 
column densities are averaged or inferred.  The data from 
Rosenberg \& Schneider (2003) can be used
to find averaged column densities for their 
sample out
to a radius where $N_{HI}=2 \times 10^{20}$ cm$^{-2}$, but their plotted HI
masses and disc areas do not tell us about the properties
of galaxies whose central column densities are about this value or lower.  
Based on their fit to the points in their first figure with a slope of one, their
characteristic average column density would be $\langle N_{HI}\rangle^* = 8.26\times  
10^{20}$ cm$^{-2}$ or $10^{20.92}$ cm$^{-2}$, which would correspond to 
$N_{max}^*= 10^{21.7}$  cm$^{-2}$
according to equation (2).  However, points scattered within about half an order
of magnitude of this value would suggest that galaxies exist that have $N_{max}$
values well below $10^{21}$ cm$^{-2}$.  Many such objects are likely to have small
HI radii above $2 \times 10^{20}$ cm$^{-2}$, and thus need to be detected in a 
deeper survey such as that of Minchin et al.~(2003).

Minchin et al.~(2003) report a characteristic $N_{HI}^{o*} = 10^{20.6}$ cm$^{-2}$ 
for the 
HIDEEP sample.  
If we assume the $N_{max}^*$ value found above, then Equation (3)
would give $N_{min}^*=10^{21.6}$ cm$^{-2}$, which would not allow for the detection
of objects with very low, or even moderate, $N_{HI}^{o*}$. This $N_{min}^*$ value 
is probably too high to make sense or
allow for the detection of many objects.  However the HIDEEP survey is likely to detect
objects with lower typical $N_{max}$ values than those of Rosenberg \& Schneider (2003).
Suppose we assume that the radii used by Minchin et al.~(2003), who use five times
the effective optical radii, are equivalent to those where $N_{min}\sim 
10^{20}$ cm$^{-2}$ as is typical of the 
galaxies studied in Salpeter and Hoffman (1996).  In this case, Equation (3)
would give two solutions: $N_{max}^*=10^{21.9}$ cm$^{-2}$ and $N_{max}^*=
10^{20.3}$ cm$^{-2}$, as seen in  Fig.~\ref{nhinfvsnmax}.  
The larger value would be unusually high compared to what
is seen in galaxies with known HI profiles or damped absorbers, although the
curve in Fig.~\ref{nhinfvsnmax} is not very steep for high $N_{max}$ values.  
On the other hand, 
objects having the lower $N_{max}$ value would likely have small radii
so
that the radii could be estimated less carefully than for larger objects.
It is thus likely that objects with a wide range of $N_{max}$ values, including
lower ones, are being
detected by HIDEEP.

A larger radius, where $N_{min}<10^{20}$ cm$^{-2}$, would be
needed to calculate $N_{HI}^{o}$ below $10^{20}$ cm$^{-2}$.
However the limiting parameter for detecting clouds and galaxies in HI
surveys is flux, rather than column density.
Thus a more interesting question, as opposed to understanding
the accuracy of galaxy radii, is whether
objects having low values of $N_{max}$ and sizes or masses similar to 
those of known galaxies, could be detected in a survey having some limiting 
flux.  We discuss this issue further using simulated galaxies in Sec. 4.

Note that column density profiles may fall off more slowly than exponentials
in the outer parts of galaxies (Hoffman et al. 1993; see discussion in Linder et
al.~2003).  Furthermore, exponential profiles are not always well-behaved
in the centres of galaxies 
where there may be stars or molecular gas instead
of neutral hydrogen. (Thus $N_{max}$ becomes the extrapolated central column density
assuming an exponential profile in the galaxy's centre.)
Finally, column density profiles are thought to fall off
quickly at a few $\times 10^{19}$ cm$^{-2}$ due to ionization, which was not
yet considered in Figs.~\ref{nhavgnmaxmin} and \ref{nhinfvsnmax}.
The effects of using more realistic column density profiles
will be discussed further for simulated galaxies.

\section{HI Profiles and Simulations}

Simulations are done in order to determine the HI fluxes for possible populations
of low column density objects, and to attempt to reproduce the distribution
of inferred column densities seen in the HIDEEP survey.
Samples of galaxies (or clouds, having unknown optical
properties) are simulated at $z=0$ in order to produce the
figures discussed in Section 4.
Gas in each galaxy is modelled as a slab structure in hydrostatic equilibrium,
where the gas is confined by a combination of pressure and gravity as in
Charlton, Salpeter, \& Hogan (1993) and Charlton, Salpeter \& Linder (1994).

We wish to simulate objects having a wide range of properties, 
especially those which are low in mass or column density which may be difficult
to detect due to ionization.
We assume that each galaxy has an exponential total
(neutral plus ionized) column density profile to start. Further simulations vary
the profile, for example using a power law fall-off beyond four HI disc scale
lengths, as discussed in Linder et al.~(2003).  
The central (total) column density
$N_{max}$ is assumed to have values which are
uniformly distributed between $10^{18}$ and $10^{22.2}$ cm$^{-2}$.
The higher central column density limit is
chosen as an upper limit of what is seen in detected galaxies and damped Ly$\alpha$
absorbers, while the lower range
is used in order to explore the possible existence of gas clouds which are more
difficult to detect, being somewhat below the current sensitivity of HI surveys.

Disc scale lengths are chosen so that the simulated objects obey a Schechter-type
total gas mass function, which gives rise to a detectable HI mass function
having a similar slope of -1.3 (Zwaan et al.~2003) or -1.5 (Rosenberg \& Schneider 2002).
Two main cases are illustrated repeatedly in the following section: In Case A, we 
assume that the central column densities of the objects are correlated with the total
gas masses.  Case A is motivated by HI observations of LSB galaxies which 
suggest that they have lower central column densities (de Blok et al.~1996)
and the likely existence of numerous dwarf LSB galaxies, such as in 
Sabatini et al.~(2003).
For each simulated galaxy a relationship is assumed for Case A
where $\log N_{max}=21.7+1.0\log(M_{tot}-10.0) \pm 0.5$, so that
objects having the lowest simulated $N_{max}$ values will tend
to also have the gas masses around $10^6 M_\odot$.  (We later vary 
this relationship, as the narrow range of scatter is chosen as an extreme example
to start.) 
In this case
we have a substantial population of small clouds having low $N_{max}$, some of 
which might resemble HVCs, although the detectable objects will tend to have 
high $N_{max}$.
In Case B the disc scale lengths and the 
central column densities
are uncorrelated.  
Thus $N_{max}$ is uniformly distributed and
unrelated
to the total gas mass, but the scale lengths tend to be larger for objects having lower $N_{max}$.
In this case we are simulating larger objects having low column densities, which
might resemble giant LSB galaxies (whose numbers are very uncertain)
or extended structures that give rise to Ly$\alpha$
absorption, as expected based upon double line of sight observations (for example,
Dinshaw et al.~1998;
Charlton, Churchill \& Linder 1995; Monier, Turnshek \& Hazard 1999; Fang et al.~1996).

Rotation velocities are found for each simulated galaxy using the relationships
given in Salpeter \& Hoffman (1996), where
the Tully-Fisher relationship between the velocity $V_{rot}$ and the observable
HI radius $R$ is found to be $V_{rot}/80.51~\mbox{km s}^{-1} =(R/12.3~\mbox{kpc})^{1.38}$.
The value of $R$ is assumed to typically correspond to a radius where the 
limiting column density is $10^{20}$ cm$^{-2}$, but then this relationship
does not give us information about
the rotation velocities of massive objects having $N_{max}<10^{20}$ cm$^{-2}$.
Since galaxies having a wide range of properties are found to obey a baryonic
Tully-Fisher relation (McGaugh et al.~2000), we extrapolate the HI Tully-Fisher
relation above into a baryonic version by assuming that $R$ in the formula above
is the radius that a galaxy of equivalent (neutral plus ionized hydrogen)
mass would have if it had $N_{max}=N_{max}^*$,
where we assume $N_{max}^*=10^{21.7}$ cm$^{-2}$ as found in the previous section.
When we attempt to simulate the smallest clouds, the value of $R$ may be very
small, so that the velocity dispersion of the gas becomes more important.
A minimum value of $V_{rot}=10$ km s$^{-1}$ is thus
assumed.  Note, however, that there is a selection effect against detecting
objects having velocity widths $\ltorder 50$ km s$^{-1}$ in HI surveys, 
as discussed in 
Minchin et al.~(2003) and Lang et al.~(2003).

The vertical ionization structure of the gas is modelled
as in Linder (1998), which is similar to the model in Maloney (1993).  Inside of
some ionization radius $R_{cr}$, the gas is assumed to have a sandwich structure, where
the inner shielded layer remains neutral and has a height ($z_i$) determined by
equation (6) in Linder (1998).  The gas above height ($z_i$) and beyond the
ionization radius is assumed to be in ionization equilibrium.

The frequency- and direction-averaged ionization rate $\zeta$ is assumed
at first to be $3.035 \times 10^{-14}$ s$^{-1}$, from the
calculation of Dav\'e et al.~(1999) at $z=0$ based upon spectra from Haardt \&
Madau (1996).  The lowest measurements at redshifts $\sim 0$ tend to be
consistent with this value as discussed in Linder et al.~(2003).
However galaxies, in addition to quasars, may contribute to
the ionizing background radiation (Giallongo, Fontana, \& Madau 1997; Shull et al.~1999;
Bianchi, Cristiani, \& Kim 2001; Linder et al.~2003).   The ionizing
intensity may be stronger when 
close to a luminous
galaxy or galaxy-rich environment as shown in Linder et al.~(2003),
although the gas-rich objects which we are simulating here are not likely
to arise in the most galaxy-rich environments.
Thus simulations are also run using a larger frequency and direction-averaged
ionizing intensity 
measurement for redshifts $z<1$ of $\zeta= 1.9 \times 10^{-13}$ s$^{-1}$ 
(Scott et al.~2002).
The conversion between $\zeta$ and a one-sided flux is assumed as in Tumlinson
et al.~(1999).
The radius $R_{cr}$,  at which the disc becomes fully ionized, is found where
the neutral gas height $z_i$ becomes zero.

For each galaxy, the neutral column density profile is calculated by integrating
the neutral density $n_{HI}$ vertically through the disc, in increments
of scale length $h/10$ (or smaller when needed for a more accurate mass 
calculation).  The profiles have $N_{HI}\sim N_{tot}$ in the
regions inside of $R_{cr}$.  Just inside the radius $R_{cr}$,
the column density falls off quickly, typically from $N_{HI}\sim 3\times 10^{19}$ cm$^{-2}$
to $N_{HI}\sim 10^{17}$ cm$^{-2}$, at which point the `ionized region' of the 
disc is being mapped, and only a small fraction of the gas is neutral.
The resulting profiles can then be integrated and averaged over suitable radii
to be compared with HI observations.

In order to calculate fluxes for the simulated objects, each object is assigned a 
random inclination and a random
distance within a sphere around us having a radius of 108 Mpc, the distance at which a 
$10^{10} M_{\odot}$ galaxy
can be detected at a limiting peak HI flux of 18 mJy/beam, as in Minchin et al.~(2003).
The HI mass for each object, limited to what can be contained within a 15 arcmin beam 
centred on the galaxy,
is binned into velocity channels of 15 km s$^{-1}$ when finding a peak or 
integrated flux, which is comparable to what is done for the HIDEEP survey.

\section{Results}

\begin{figure}
\vspace{0.6cm}
\includegraphics[width=84mm]{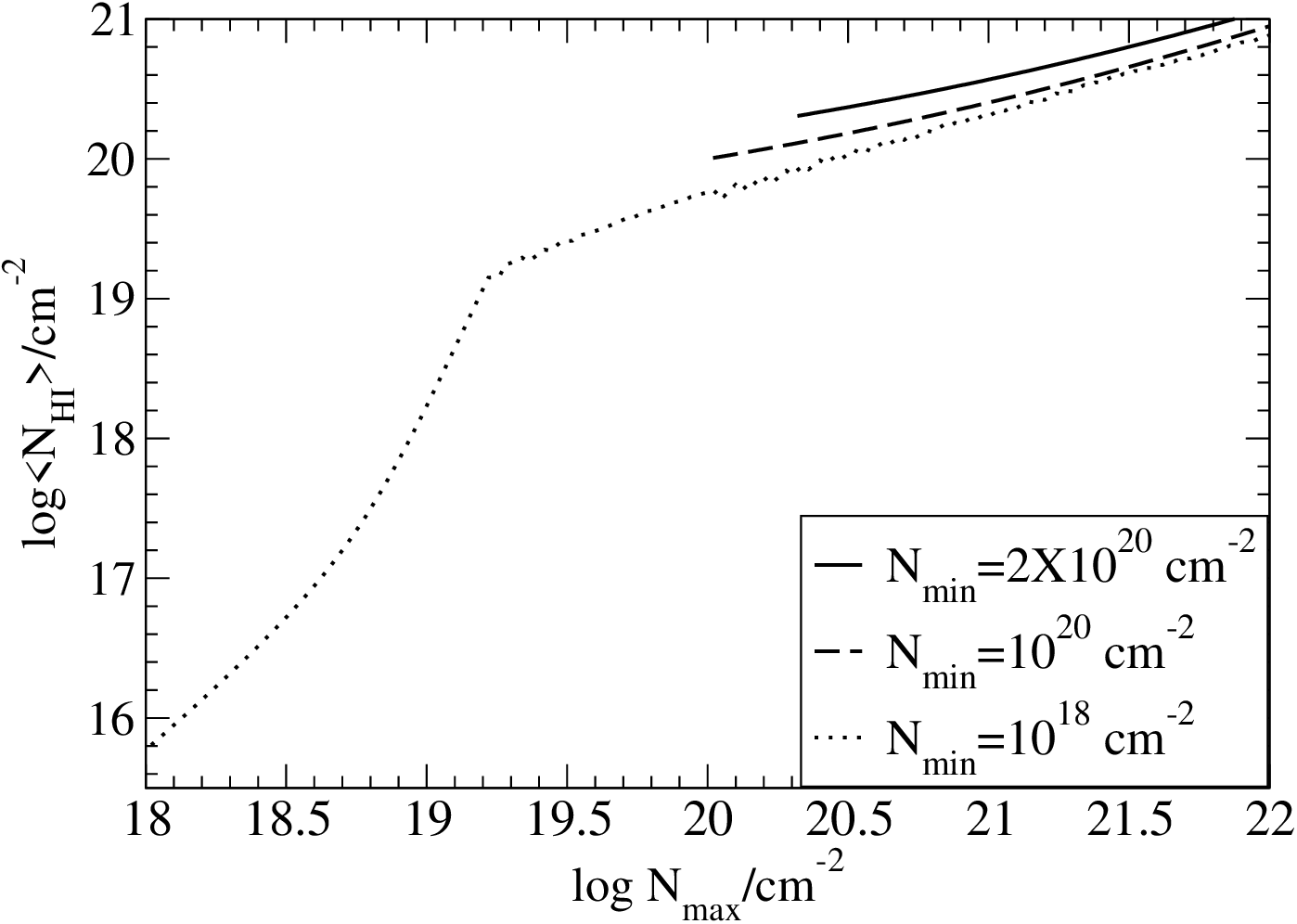}
 \caption{Averaged neutral column densities 
are plotted versus central neutral column densities ($N_{max}$) for simulated
galaxies having exponential column density profiles with $h=2$ kpc which are
exposed to an ionizing background.  
Column densities are 
found here by integrating HI masses out to $N_{min}$ and dividing by the area
within the radius of $N_{min}$, where  $N_{min}$ ranges from $2\times 10^{20}$ 
cm$^{-2}$ (top curve) to
$10^{18}$ cm$^{-2}$ (bottom), which is indistinguishable from $N_{min}=
10^{19}$ cm$^{-2}$ (for $N_{max} > 10^{19}$ cm$^{-2}$) due to the steep column
density profiles at these column densities resulting from ionization.  
The highest curve has $N_{min}$ equal to
that in Rosenberg \& Schneider (2003).
The range of 
$\log \langle N_{HI}\rangle$ in this curve shown here is fairly narrow, 
thus showing that some variation
is possible in $N_{max}$ for observed galaxies.
}
\label{nhavgvsnmaxexp}
\end{figure}

Averaged column densities are plotted, again versus central column density $N_{max}$,
for simulated galaxies, which are exposed to an ionizing background.  
For simulated galaxies $N_{max}$ is defined as the total (neutral plus ionized) 
assumed central column density, which is about equal to the observable neutral
value for $N_{max}>$ a few $\times 10^{19}$ cm$^{-2}$.  For lower $N_{max}$ values,
the neutral values could be as low as a few $\times 10^{17}$ cm$^{-2}$, although
it is difficult to determine the value accurately very close to the centres of the discs
with the model used here.  Gas with column densities $\ltorder 10^{19}$ 
cm$^{-2}$ are seen, for example as mini-HVCs (Hoffman et al.~2002).

\begin{figure}
\vspace{0.5cm}
\includegraphics[width=84mm]{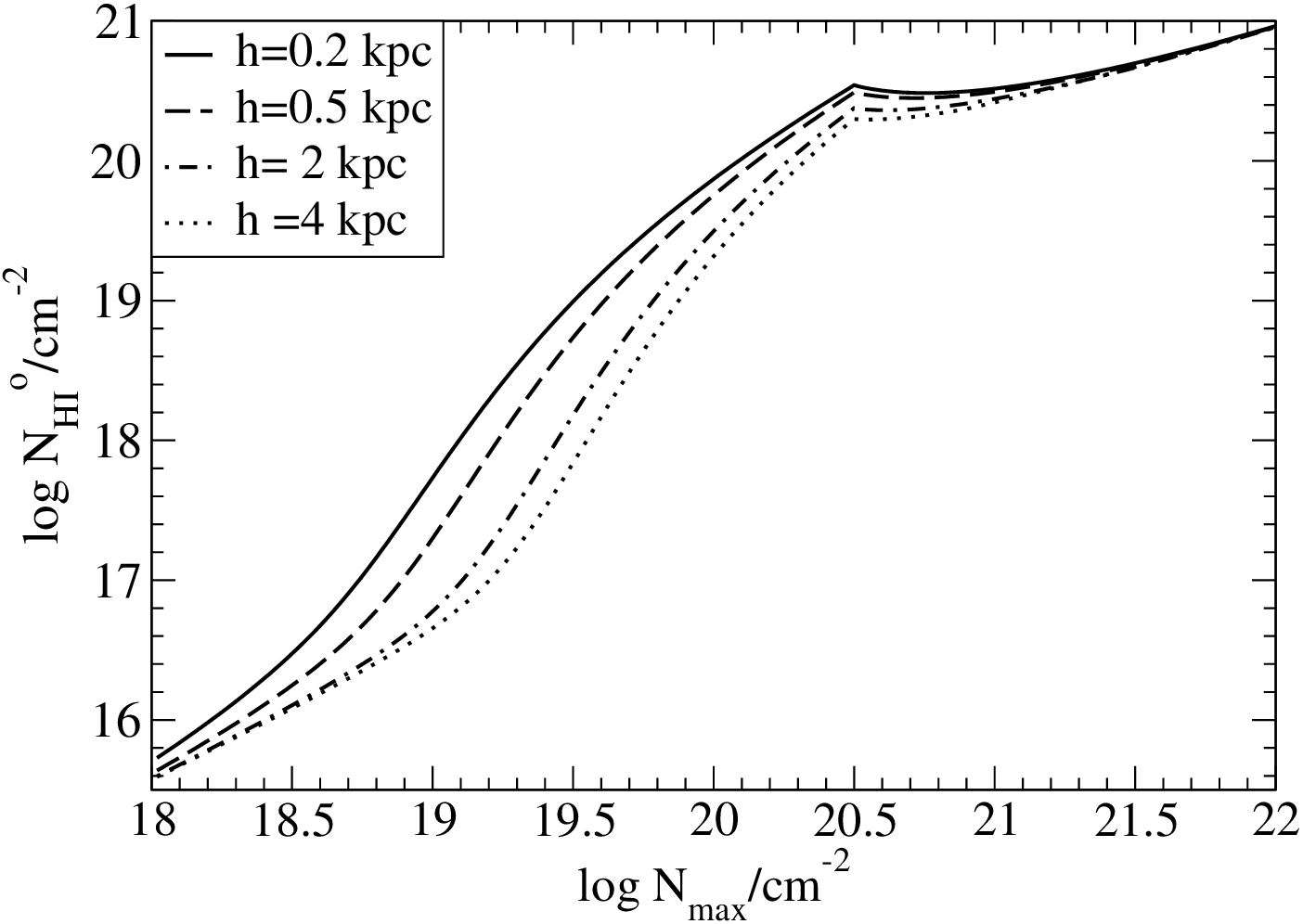}
 \caption{Inferred neutral column densities
are plotted versus central neutral column densities ($N_{max}$) for simulated
galaxies which are exposed to an ionizing background.  Column densities are
found here by dividing total HI masses by the area
within the radius of $N_{min}$, where  
 $N_{min}=10^{20}$ cm$^{-2}$ or that
$\log N_{min}=\log N_{max}-0.5$ if $\log N_{min} < 20.5$.
Curves are shown for several disc scale length values, ranging from $0.2$ to 
4 kpc.  Note that the curves are much steeper at low $N_{max}$ than the line 
shown in Fig.~\ref{nhinfvsnmax} due to ionization.
}
\label{nhinfvsnmaxexp}
\end{figure}

\begin{figure}
\vspace{0.5cm}
\includegraphics[width=84mm]{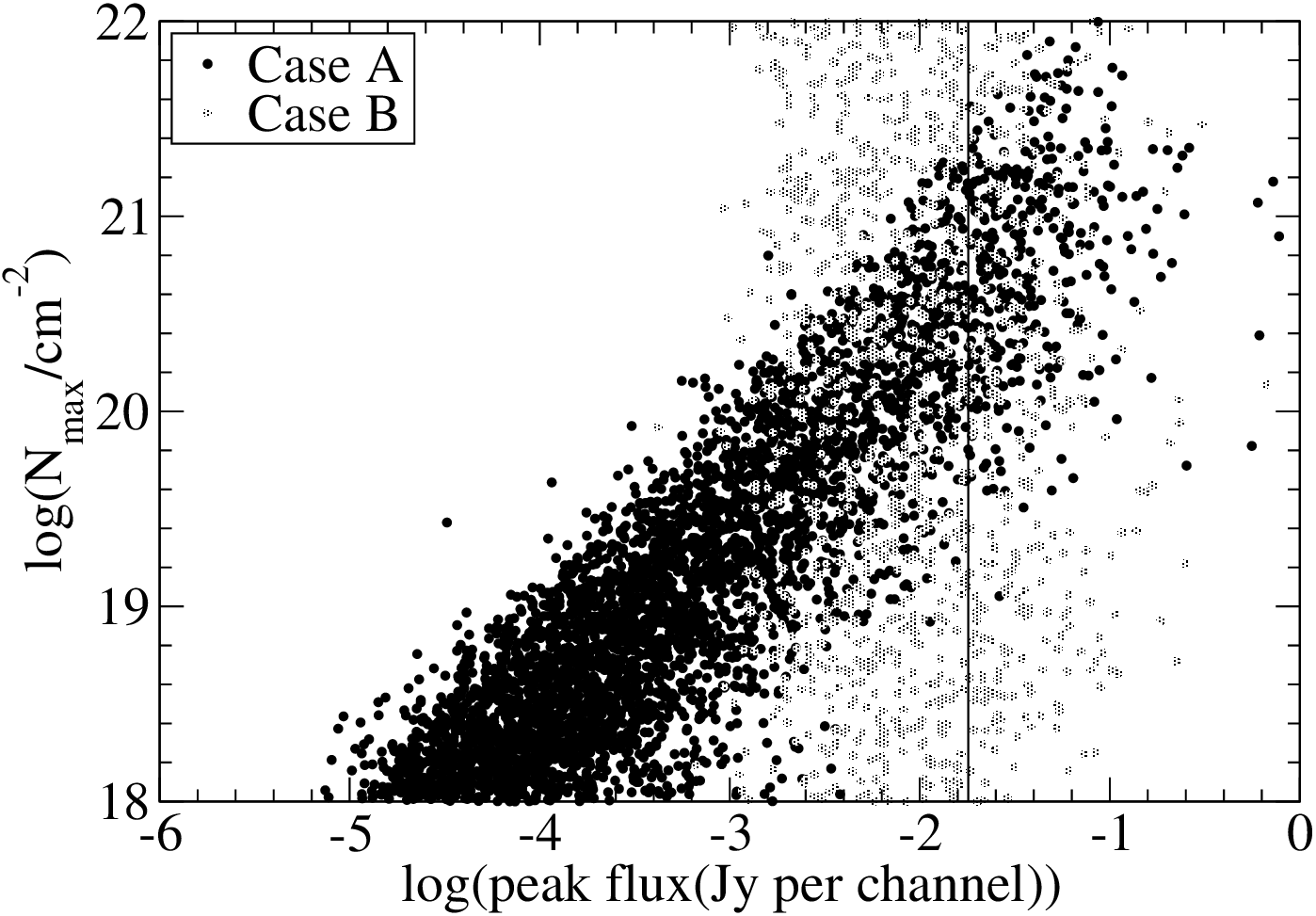}
\caption{Without an ionizing background, central column densities $N_{max}$
are plotted versus peak fluxes for simulated galaxies
having purely exponential profiles, both for a sample of objects where the
HI masses are correlated with $N_{max}$ (Case A, black circles) and for a sample
where the masses are independent
of central value $N_{max}$ (Case B, grey circles).  A line is drawn at the limiting flux of the HIDEEP
survey.  If the lower column density galaxies were unionized, as shown here,
galaxies having a wider range of central column densities would be detectable.}
\label{ntmaxvsflux}
\end{figure}

\begin{figure}
\vspace{0.5cm}
\includegraphics[width=84mm]{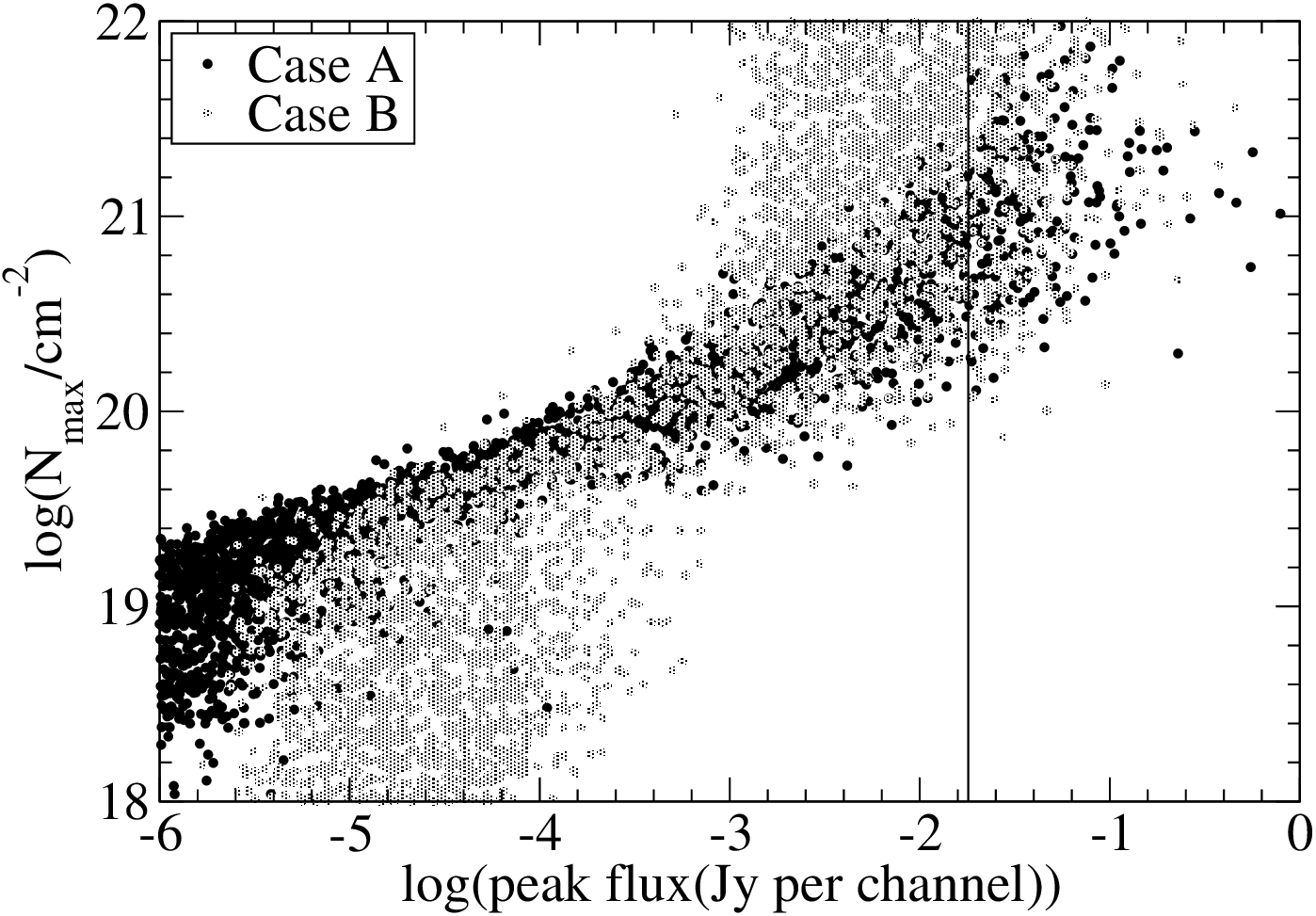}
\caption{Central column density ($N_{max}$)
values are plotted versus peak fluxes for simulated 
galaxies, for the cases shown in Fig.~\ref{ntmaxvsflux}, but now exposed to an 
ionizing background.
Few points having $N_{max}$
below $10^{20}$ cm$^{-2}$ can be detected below the limiting flux of the 
HIDEEP survey.  
}
\label{nmaxvsflux}
\end{figure}

\begin{figure}
\vspace{0.5cm}
\includegraphics[width=84mm]{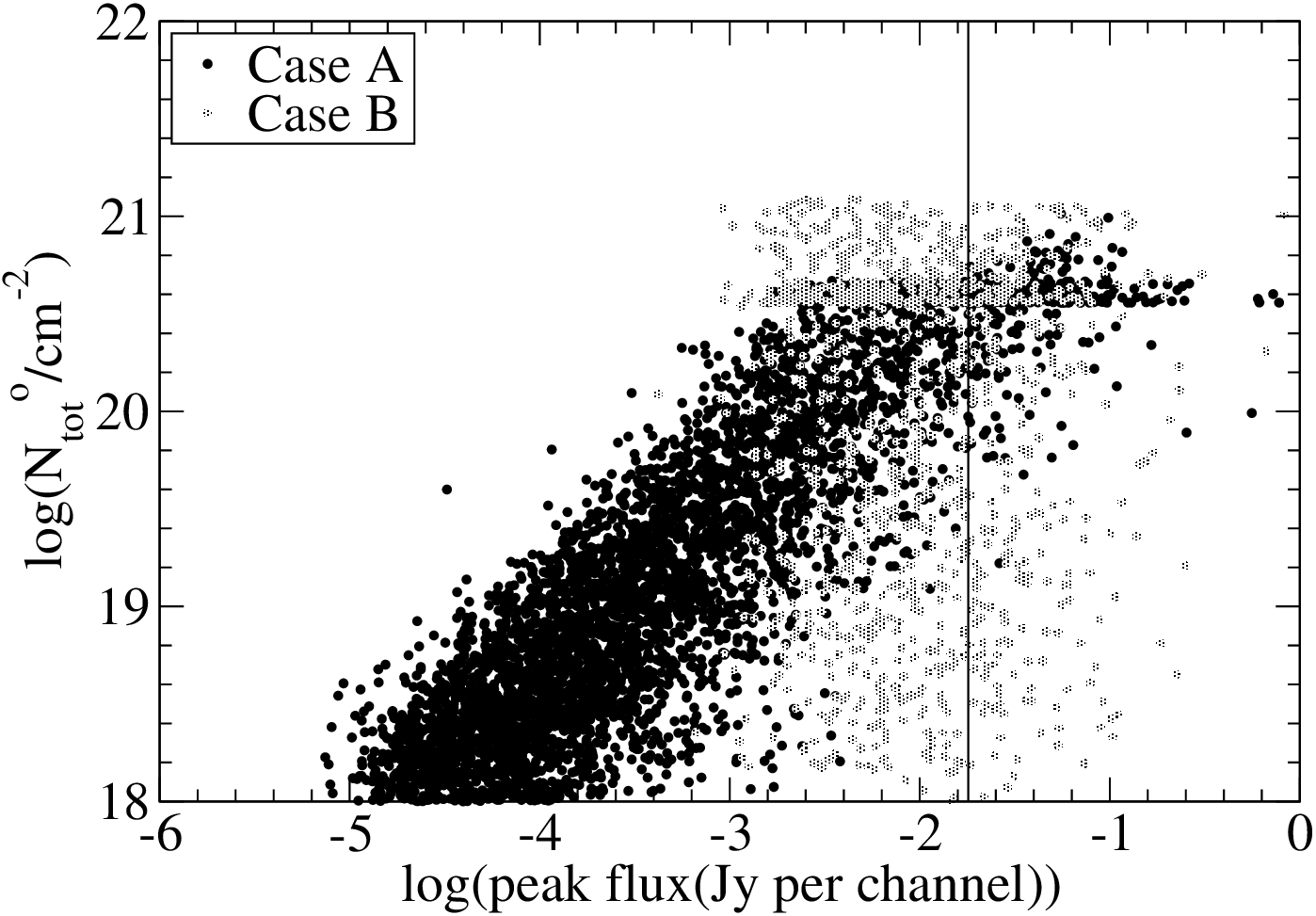}
\caption{Inferred column density values ($N_{tot}^o$), without an ionizing background
are plotted versus peak fluxes for simulated galaxies, again for the cases
described in Fig.~\ref{ntmaxvsflux}, assuming radii where $\log N_{min}=20$
or $\log N_{min}=\log N_{max}-0.5$ if $\log N_{max}<20.5$.  The majority of
the points, which are above the HIDEEP limiting flux, are seen at $10^{20.6}$ 
cm$^{-2}$.  Some lower $N_{tot}^o$ values could be detected, in this 
hypothetical unionized case, if the radii of these objects were correctly
estimated.
}
\label{ntinfvsflux}
\end{figure}

\begin{figure}
\vspace{0.5cm}
\includegraphics[width=84mm]{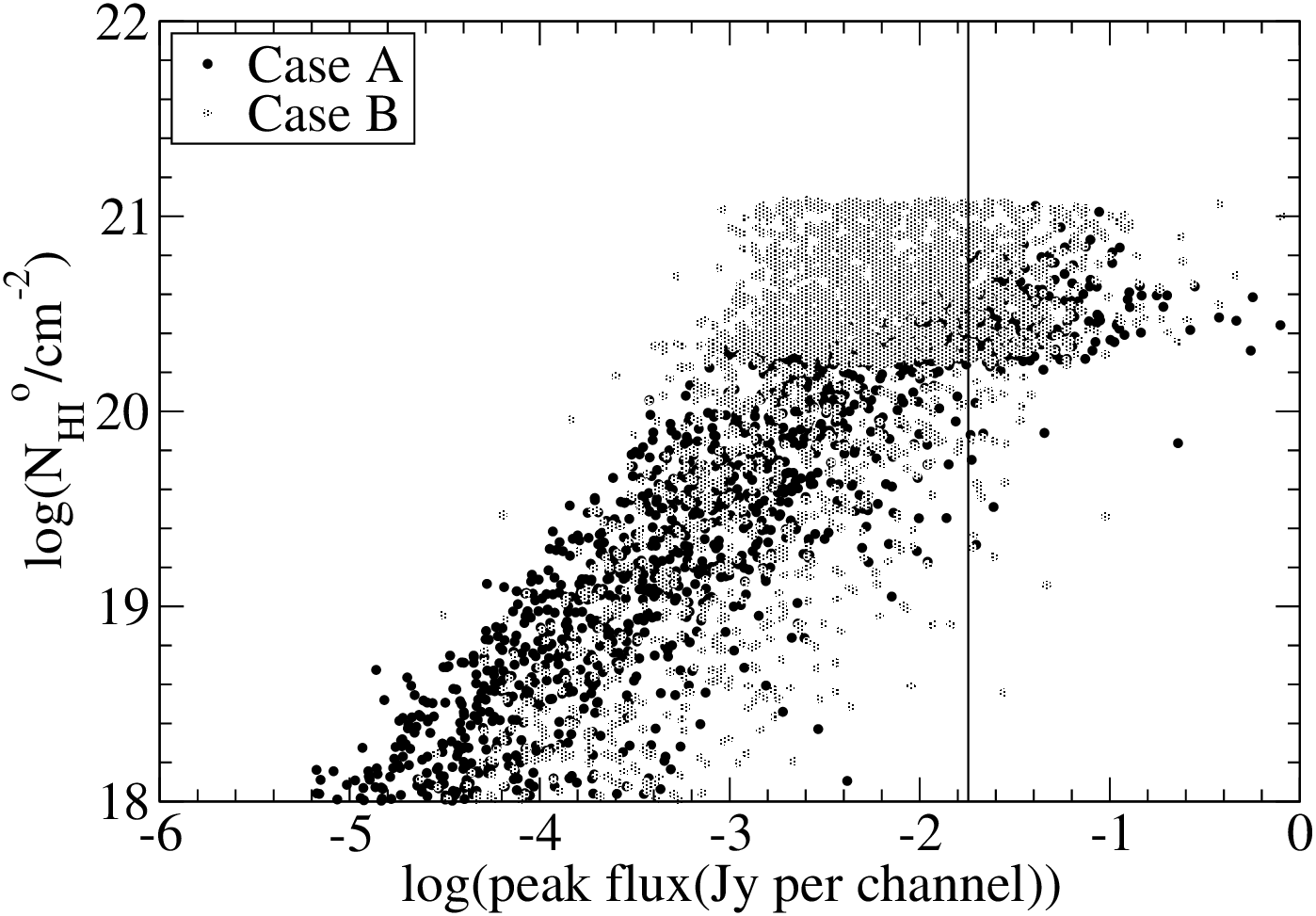}
\caption{Inferred column density ($N_{HI}^o$)
values are plotted versus peak fluxes for simulated galaxies,
for the two cases described in Fig.~\ref{ntmaxvsflux}  Here the gas has been exposed
to an ionizing background, and $N_{min}$ values are assumed as in 
Fig.~\ref{ntinfvsflux}  Above the limiting HIDEEP flux, only a
few points might be detected having $N_{max}
\ltorder 10^{20}$ cm$^{-2}$.
}
\label{nhinfvsflux}
\end{figure}

Purely exponential total column density profiles
are assumed for the objects simulated in Figs.~\ref{nhavgvsnmaxexp} through 
\ref{nhinfvsflux}.  In 
Fig.~\ref{nhavgvsnmaxexp}, the
column densities are averaged, so that the mass contained within
a radius where $N_{HI}=N_{min}$ is divided by the area within this radius.
Thus the right side of the plot looks similar to Fig.~\ref{nhavgnmaxmin}, but the
left side shows where the averaged column densities become lower when most of the
mass in a galaxy or HI cloud is close to the ionization edge.  

Samples of 200 objects, having uniformly distributed values of $\log N_{max}$,
are simulated in Figs.~\ref{nhavgvsnmaxexp} and \ref{nhinfvsnmaxexp}.
We simulate only galaxies having $h=2$ kpc in Fig.~\ref{nhavgvsnmaxexp}, as
the curve will otherwise become
widened in the steeper parts ($N_{max} \ltorder 10^{20}$ cm$^{-2}$)
 due to variations in disc scale lengths.
The top curve shows objects observed to the same limits as in 
Rosenberg \& Schneider (2003).  It can be seen that less than one order of magnitude
of variation in $\langle N_{HI}\rangle$ corresponds to more than two orders of magnitude
in possible $N_{max}$ values.

In Fig.~\ref{nhinfvsnmaxexp} we plot inferred column densities ($N_{HI}^o$), 
also versus central column 
density $N_{max}$, for simulated objects
which are exposed to an
ionizing background.  Inferred column densities are found by dividing the total
HI mass of a galaxy or cloud by an area within some estimated HI radius.
One might use a radius where ($N_{min}\sim 10^{20}$ cm$^{-2}$) which would be typical
of the galaxies discussed in Salpeter and Hoffman (1996), but it would be necessary
to use a radius corresponding to a smaller $N_{min}$ to find a smaller $N_{HI}^o$.
Thus we assume here (and for further calculations of $N_{HI}^o$)
 that $N_{min}=10^{20}$ cm$^{-2}$ or that 
$\log N_{min}=\log N_{max}-0.5$ if $\log N_{min} < 20.5$.
In Fig.~\ref{nhinfvsnmaxexp} we show several scale length values between 0.2
and 4 kpc.  The upper parts of the curves (having $\log N_{max} > 20.5$) are similar
to the central curve in Fig.~\ref{nhinfvsnmax}, although any variation with $h$ happens
only with an ionizing background.  Here the lower curves, where $\log N_{min}=\log N_{max}-0.5$
are steepened due to ionization, when compared to the line with slope 0.745 without ionization.

It can be seen in Figs.~\ref{nhavgvsnmaxexp} and \ref{nhinfvsnmaxexp} that the 
averaged and inferred column densities both fall off quickly when $N_{max}$ goes
below some value $\ltorder 10^{20}$ cm$^{-2}$, though the $N_{HI}^o$ or $\langle N_{HI}\rangle$ 
value where
this happens depends upon the assumed value of $N_{min}$.
The steepening of the dotted curve in Fig.~\ref{nhavgvsnmaxexp} is 
a result of galaxy discs having ionization edges at a few $\times 10^{19}$ 
cm$^{-2}$.  
For the inferred column densities shown in Fig.~\ref{nhinfvsnmaxexp} the steepening happens
in part because of the radius withinin which the column 
density is averaged, as it is for the dotted line shown in Fig.~\ref{nhinfvsnmax}. However,
the line in Fig. 2 has
a slope of 0.745, whereas in Fig.\ref{nhinfvsnmaxexp},
$N_{HI}^o$ changes by about four orders of magnitude when 
 $N_{max}$ changes by two
orders of magnitude, which is as steep as a line with a slope of $2$, again resulting
from ionization.
For objects
having low $N_{max}$, the area of the disc which has a 
high column density becomes small, so that the HI fluxes for these objects are
also likely to be small.

While $N_{HI}^o$ values may difficult to estimate, we ultimately want to know which 
central column density 
$N_{max}$ values can be detected in a survey having some limiting
flux.  In Figs.~\ref{ntmaxvsflux} through \ref{nhinfvsflux}, we plot peak fluxes for simulated
galaxies, which can be compared with the HIDEEP limiting value of 18 mJy per
velocity channel, 
where the velocity resolution is 18 km/s and the channel separation is 13.2 km/s.
corresponding to the vertical line in each Figure.
Samples of 5000 objects, having
total gas masses between $10^6$ and $10^{11}$ $M_{\odot}$ which obey a total gas mass
function with a slope of -1.3,
are simulated for each case.

In Fig.~\ref{ntmaxvsflux}, central column density values
are plotted versus 
peak flux values which would be produced by completely unionized gas.
More objects have lower fluxes than higher ones at any given $N_{max}$, as there
are more objects which are at larger distances, and more having lower masses.
In Case A (black
circles), where the simulated masses are correlated with $N_{max}$, and thus low $N_{max}$ 
objects tend to have smaller scale lengths, an observer would not expect to detect many
objects having $N_{max}\ltorder 10^{20}$ cm$^{-2}$, even without considering
the effects of ionization.  Yet if the objects with low $N_{max}$ are as 
massive as those having high $N_{max}$ (Case B, grey circles), then numerous 
objects having low $N_{max}$ would still be seen above a reasonable flux limit.
The fluxes for objects having lower $N_{max}$ values would only be lower if 
a substantial fraction of the mass is outside the beam, as seen for 
$N_{max}\ltorder 10^{19}$ cm$^{-2}$, or if some of the gas is ionized, as
seen in Fig.~\ref{nmaxvsflux}.
In Fig.~\ref{nmaxvsflux}, values of $N_{max}$ are plotted versus peak flux
for each object, now including the effects of ionization. 
At a limiting flux above that in the HIDEEP survey, few objects 
could be detected having $N_{max}>10^{20}$ cm$^{-2}$ (although the detectable inferred column
densities could be lower than this value, as seen in Fig. ~\ref{nhinfvsflux}).
It can be seen that an observer would need to look at least two orders of magnitude
fainter than the HIDEEP limit to detect ionized objects having 
$N_{max}\ltorder 10^{20}$ cm$^{-2}$ in Case B, or possibly more if low column 
density objects are less massive.
The fluxes shown, for example, for ionized objects here, may be lower limits
if the gas far from galaxies is clumpy.

In Figs.~\ref{ntinfvsflux} and \ref{nhinfvsflux}, we attempt to calculate 
inferred column densities, which are plotted again versus peak flux, and
can be compared with the $N_{HI}^o$ values for the HIDEEP survey.  Again we find
$N_{HI}^o$ values within radii where  $\log N_{min}=20$ or $\log N_{min}=
\log N_{max}-0.5$ if $\log N_{max}<20.5$, which makes sense, as all the 
objects detected in
HIDEEP were found to have possible optical counterparts, and are thus assumed
to have measurable HI radii.  In Fig.~\ref{ntinfvsflux} we show the 
total inferred column densities ($N_{tot}^o$), or the neutral gas that would
be seen if there were no ionizing background.  A flat distribution of $N_{max}$
values has been assumed all along for Case B, while even more objects
with low $N_{max}$ are simulated in Case A,
yet the majority of the points which are above
the limiting flux are seen at $N_{tot}^o \sim 10^{20.6}$
cm$^{-2}$.  Thus the $N_{HI}^o$ values found in HIDEEP, where the distribution
is peaked at $10^{20.65}$ cm$^{-2}$, should have a strongly
peaked distribution simply as a result of averaging exponential profiles over 
a radius where $N_{min}=10^{20}$ cm$^{-2}$.  Yet numerous lower $N_{tot}^o$
would still be seen, especially for Case B.

\begin{figure}
\vspace{0.7cm}
\includegraphics[width=84mm]{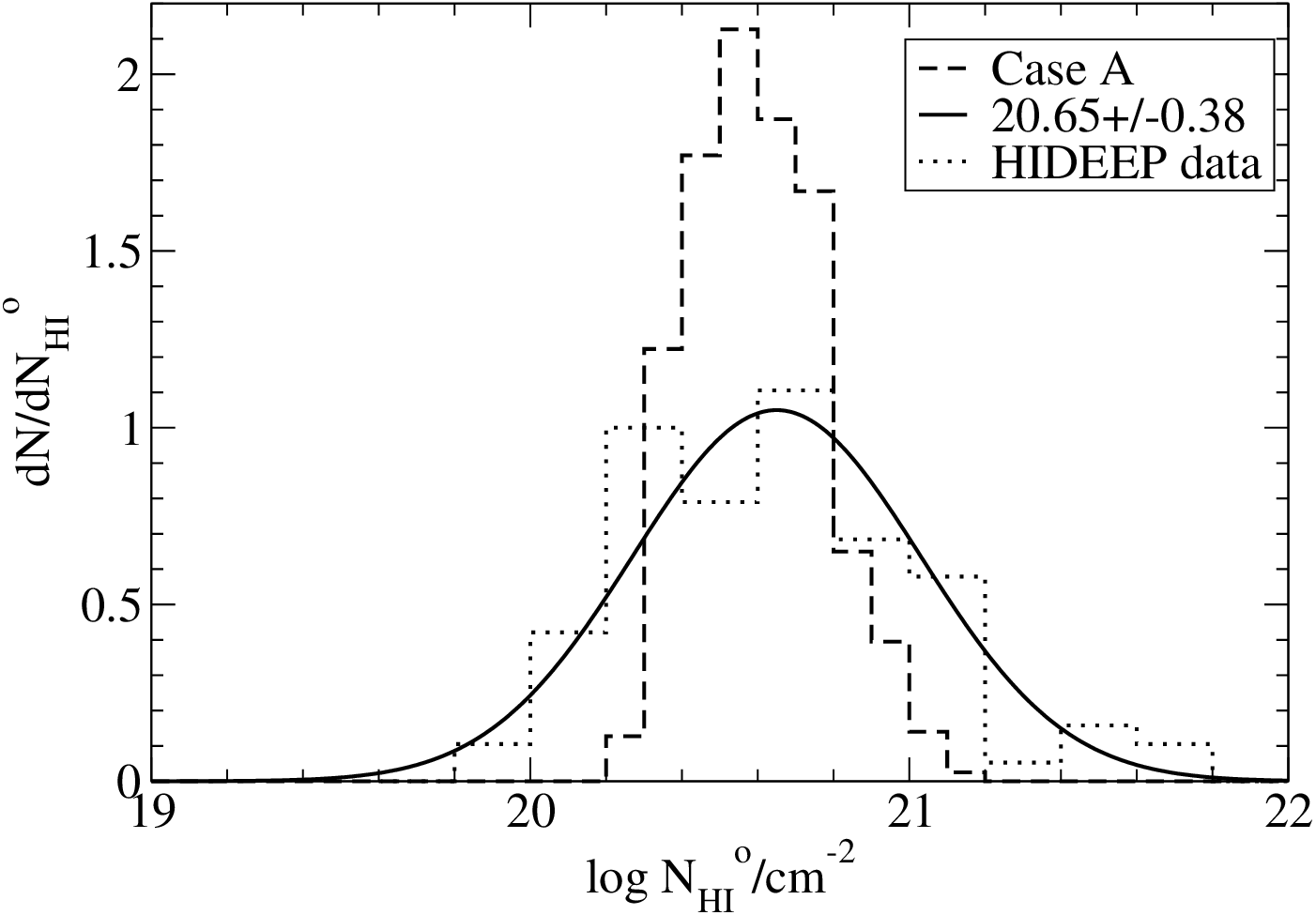}
\caption{Distributions are shown for simulated inferred column density ($N_{HI}^o$)
values detected above the limiting flux of HIDEEP, for Case A, where the
gas mass for each galaxy is closely
related to $N_{max}$ (dashed line).
Also shown is the Gaussian curve fitted to the HIDEEP sample
(solid line) and the binned HIDEEP data (dotted line).
}
\label{nhinfdista}
\end{figure}

\begin{figure}
\vspace{0.7cm}
\includegraphics[width=84mm]{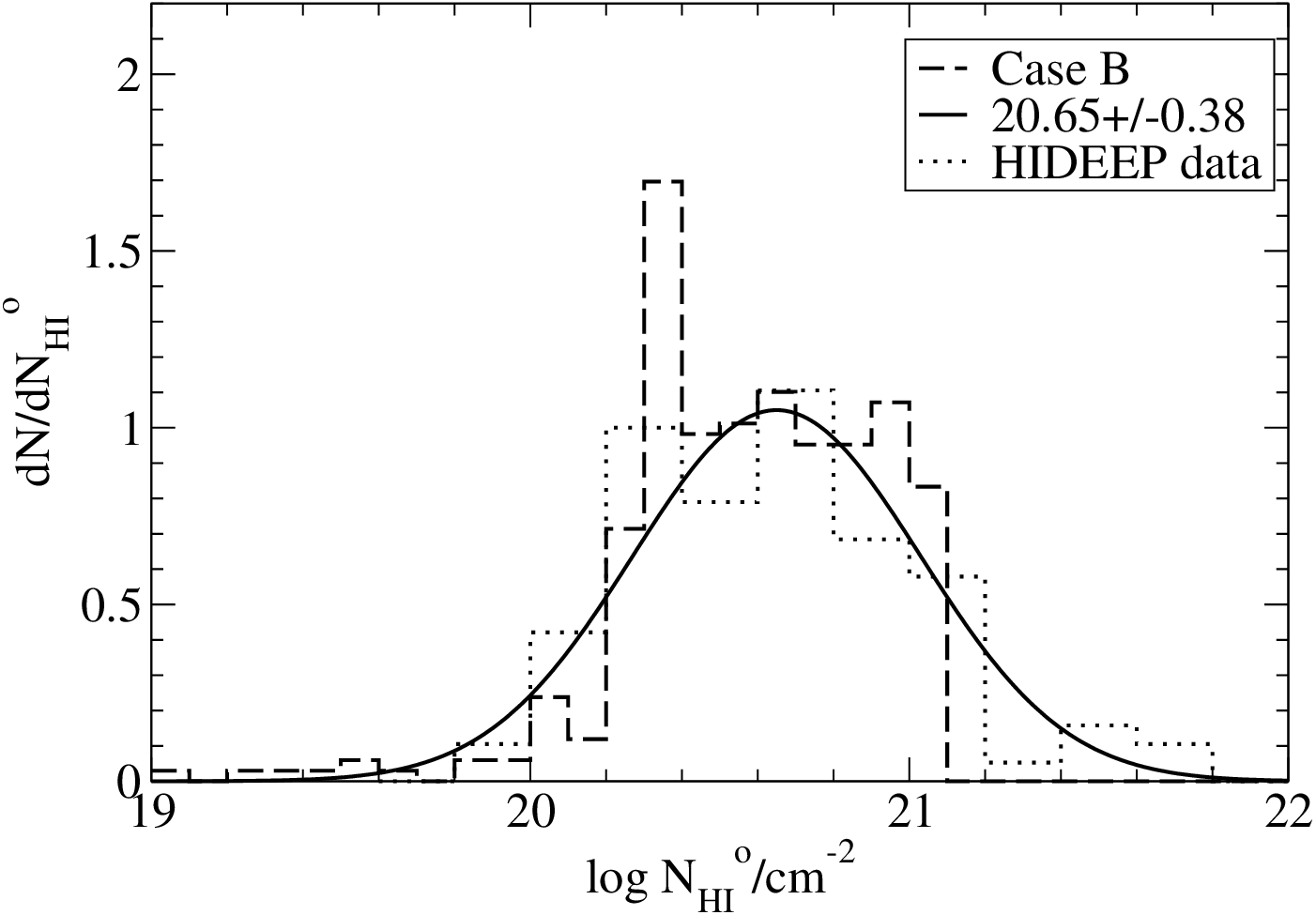}
\caption{Distributions are shown for simulated $N_{HI}^o$
values detected above the limiting flux of HIDEEP, for Case
B, where the total gas mass is independent of
$N_{max}$ (dashed line).  Also shown is the Gaussian curve fitted to the HIDEEP sample
(solid line) and the binned HIDEEP data (dotted line).
}
\label{nhinfdistb}
\end{figure}

The $N_{HI}^o$ values, as seen when the gas is exposed to an ionizing 
background, are shown in Fig.~\ref{nhinfvsflux}.  Here only a few galaxies
or clouds having $N_{HI}^o<10^{20}$ cm$^{-2}$ are seen above the HIDEEP flux 
limit.  
In either of the cases, it should be possible to detect more objects having
low $N_{HI}^o$ in a survey which is just slightly deeper than HIDEEP, if 
the radii of the objects can be estimated accurately enough.
For Case A the objects shown appear to have a similar relationship
between peak flux and column density as in the unionized plot, but the main
difference is that 
more of the objects have inferred column densities that are below the
minimum value shown on the plot.
Here the distribution of detectable $N_{HI}^o$ values appears to be
somewhat less strongly peaked compared to the unionized cases, yet the 
binned points, which are simulated like those above the HIDEEP
flux limit in  Fig.~\ref{nhinfvsflux}, do not look very different from the
HIDEEP distribution, as shown in 
Figs.~\ref{nhinfdista} and \ref{nhinfdistb}. 

In Figs.~\ref{nhinfdista} and \ref{nhinfdistb} 
we show the inferred column density distributions
for Cases A and B respectively, for samples of $1000$ objects having peak fluxes
$>18$ mJy and velocity widths $>40$ km/s.  Also shown is the Gaussian curve 
fitted to
the HIDEEP galaxies, having a mean of 20.65 and a scatter of 0.38,
as in Minchin et al.~(2003) and the binned data from Minchin et al.~(2003).  
Neither histogram is very different from the 
Gaussian or data set, assuming some measurement uncertainties.
Case A appears to be somewhat strongly peaked
here, although the observational uncertainties, mostly in measuring the
area of the galaxies, are not yet shown here.  Case B appears to be 
more broad, and similar to 
the plotted Gaussian.  The peak feature at $10^{20.3}$
cm$^{-2}$ is also seen in the Minchin et al.~(2003) 
data.  This feature is a result of many galaxies with high $N_{max}$ 
having their column densities averaged over similar $N_{min}$, and 
confirms that similar galaxy radii are being used 
to find the inferred column densities for the majority of the galaxies
for the observations and for the simulations done here.  The Case A 
model looks more realistic at the high column density end of the 
distribution, although we do not model high column density galaxies 
carefully given that more of their gas may be converted into
stars or ionized by these stars.  Neither distribution is very different
from the Gaussian curve, although the most realistic scenario 
is likely to have an intermediate behaviour between Cases A and B.  There
is likely to be some relationship between the gas masses and central column
densities of galaxies, but the scatter may be larger than
we have assumed in Case A.   For example, we increase the amount of 
scatter for $N_{max}$ from 0.5 to 1.5 orders of magnitude in Case C.

In Case B (Fig.~\ref{nhinfdistb}) there are a few objects seen
which have sufficiently high fluxes to be detected,
but column densities below anything
detected by HIDEEP.  (These points are not seen for Case A, only because
sufficiently low column densities are assumed to arise only in very low mass
galaxies which are all below the HIDEEP flux limit.)  Such points may
have simply not yet been detected by HIDEEP due to the small number of objects detected,
compared to the 1000 objects simulated here.
Also, these objects may have small, and thus uncertain radii.
Very few
HI clouds are thought to have no optical counterparts (Davies et al.~2004),
and there are no clouds detected which have no optical counterparts in 
HIDEEP.
This could happen because the gas is actually more clumpy than 
we have modelled here, so that star formation occurs in small regions
within these clouds.  The isolated HI cloud detected by Giovanelli \& Haynes (1989),
for example, was later found by many observers to contain a small dwarf galaxy.
The radii for such objects are thus likely to be underestimated, so that
their inferred column densities will be higher than what we find here.

The inferred column density distributions for the 
unionized cases are surprisingly indistinguishable from those
shown in Figs.~\ref{nhinfdista} and \ref{nhinfdistb}, when assuming that galaxies having
velocity widths $<40$ km/s are not detectable, although these objects
were not removed from Figs. ~\ref{ntmaxvsflux} through 
\ref{nhinfvsflux}.  Thus velocity width related selection effects could
be as important as ionization 
in determining the shape of the distribution of inferred column
densities. Objects which could be ionized to the point of being
undetectable have low masses and thus low velocity widths in Case A.
In Case B the distribution of $N_{HI}^o$ is the same for detected and
undetected objects.  However the neutral and ionized column density 
distributions are not the same in the intermediate Case C, where 
$\log N_{max}=21.7+1.0\log(M_{tot}-10.0) \pm 1.5$, as some galaxies
having detectable velocity widths are affected by ionization, yet there
is some variation with galaxy mass.
In Case C the $N_{tot}^o$ distribution is very strongly peaked, yet the
$N_{HI}^o$ distribution is somewhat broad, and similar to that seen
for Case B.

\begin{figure}
\vspace{0.6cm}
\includegraphics[width=84mm]{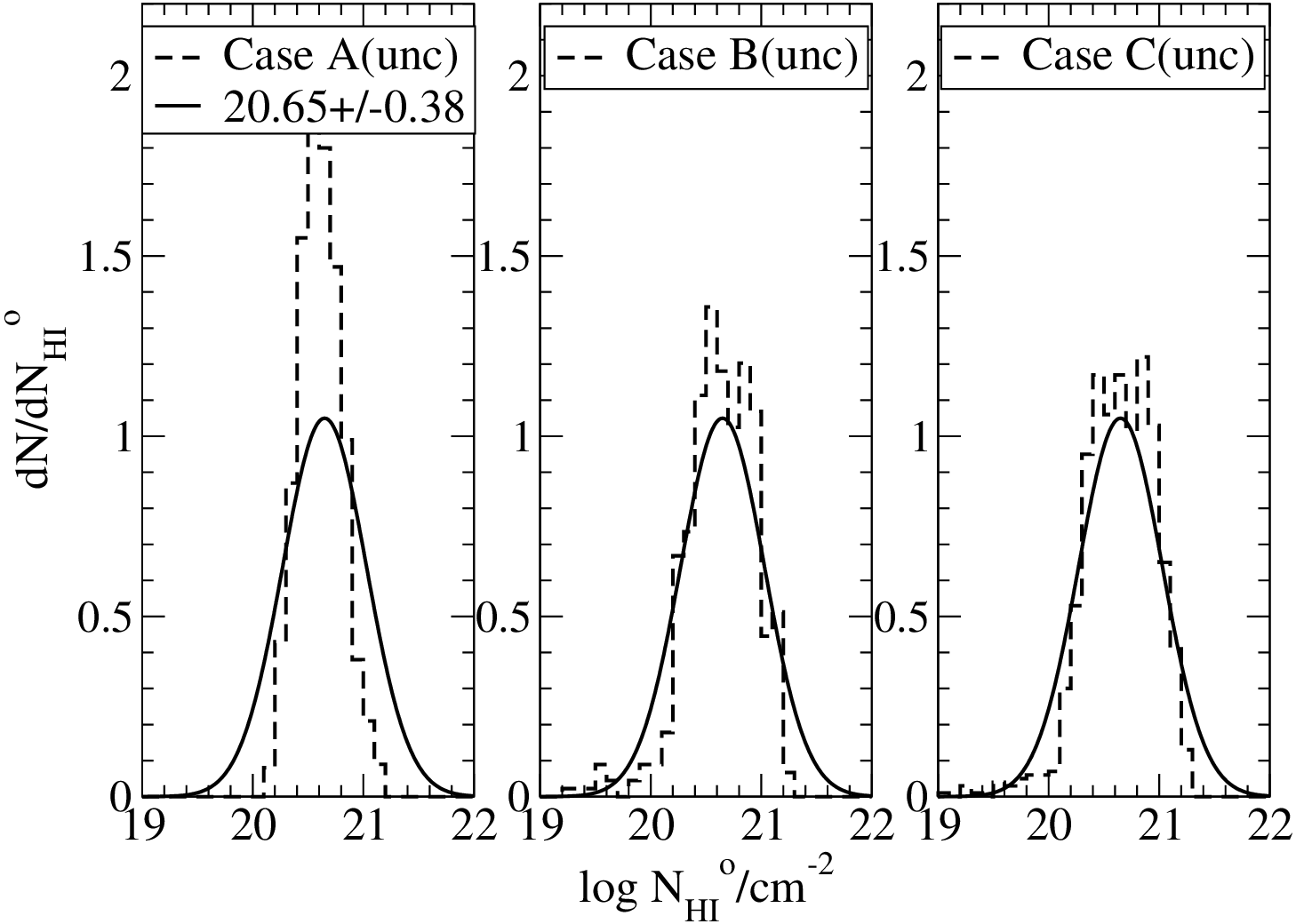}
\caption{Distributions of simulated $N_{HI}^o$ (dashed lines)
are shown with uncertainty in
the measured galaxy radii included for Cases A (left),
B (centre), and C (right).
Also shown is the Gaussian curve fitted to the HIDEEP sample in each
frame
(solid line).
}
\label{nhinfdistuc}
\end{figure}

The biggest source of uncertainty in finding inferred column densities is
in measuring the galaxy radii.
When an uncertainty of $40\%$ in the radii is included, as shown in
Fig.~\ref{nhinfdistuc}, Case A is still too strongly peaked, while 
Cases B and C are more similar to each other and to the Gaussian curve fitted
to the HIDEEP data.  Fluctuations in the ionizing background radiation
are also likely to broaden the distributions somewhat, but
only by increasing the number of galaxies with high $N_{HI}^o$ as there
are few locations in space (Linder et al.~2003)
where the ionizing background is as low as the value assumed here (Haardt \&
Madau 1996).

A summary of the 
simulations is given in Table~\ref{table1}, where the mean and scatter
values for the $N_{HI}^o$ distributions are listed.  Case C
is intended to be intermediate between the Cases A and B shown previously, as
there is some correlation between $N_{max}$ and the gas masses, but a larger
scatter so that $\log N_{max}=21.7+1.0\log(M_{tot}-10.0) \pm 1.5$. 

In an attempt to put some constraints on the properties of the objects detected
in the HIDEEP survey, some further variations were made of the model parameters.
The slope in the $\log N_{max}$--$\log M_{tot}$ relationship was increased from
$1.0$ to $1.2$, which resulted in a strongly peaked inferred column density
distribution similar to that
for Case A.
In further simulations we assume that 
$\log N_{max}=21.7+1.0\log(M_{tot}-10.0) \pm 1.5$ as in Case C.
Flattening the central parts of the galaxy column density profiles 
(to a value between $10^{20.5}$ and $10^{21.5}$ cm$^{-2}$) also 
gave rise to a strongly peaked distribution as in Case A.
We also simulated a stronger ionizing background,
based upon the $z<1$ measurements of Scott et al.~(2002), and did a 
simulation using a steeper slope of $-1.5$ for the HI mass function, as reported by Rosenberg \&
Schneider (2002).   In both of these cases the inferred column distribution is similar
to that in Case C.  Furthermore we varied the characteristic 
$N_{max}$ value, so that $\log N_{max}=22.2+1.0\log(M_{tot}-10.0) \pm 1.5$ while
using flattened central column density profiles in order to allow for
higher extrapolated $N_{max}$ values as suggested in Bowen, Blades, \& Pettini (1996).
In this case the $ N_{HI}^o$
distribution is somewhat strongly peaked.
A further simulation was done where the exponential column density profiles
were flattened to a power
law with a slope of $-4$ beyond four HI scale lengths,
in which case we see a slightly excessive number of high column density
galaxies.  
Combining the slow fall-off in the outer parts with a reduced
central column densities would likely give rise to a more realistic
number of high column density galaxies.  
For any of the parameter variations, the distribution is
not very different from the Gaussian curve fitted to the HIDEEP points, given
the uncertainty in the actual relationship between gas masses and central 
column densities. 

\begin{table}
\caption{Inferred Column Densities and Lyman Limit Absorber Counts}\label{table1}
 \begin{tabular}{@{}cccc}
  \hline
  Case
        &$\langle\log N_{HI}^o\rangle$ & $\sigma(\log N_{HI}^o)$ &$(dN/dz)_{0,LL}$\\ 
  \hline
 A&20.65&0.22&2.61\\
 B&20.62&0.30&0.93\\
 C&20.63&0.30&1.22\\
Au&20.61&0.20&\\
Bu&20.63&0.33&\\
Cu&20.61&0.32&\\
  \hline
 \end{tabular}

\medskip
Mean inferred column densities, scatter for the inferred column density
distributions, and Lyman limit absorber counts are 
shown for the simulations as summarized here: \newline  Case A, where gas masses are
related to $N_{max}$ such that $\log N_{max}=21.7+1.0\log(M_{tot}-10.0)\pm 0.5$; \newline Case B, where gas masses are independent of 
$N_{max}$; \newline Case C, where
$\log N_{max}=21.7+1.0\log(M_{tot}-10.0)\pm 1.5$; \newline
Cases Au, Bu, and Cu, versions including uncertainties in galaxy radii.
\end{table}

\section{Lyman limit absorber counts}
Lyman limit absorber counts, arising from quasar lines of sight through
gas having $10^{17.2}<N_{HI}<10^{20.3}$ cm$^{-2}$ provide further constraints
on the numbers and properties of undetected objects containing low column 
density gas.  Estimates have been made for the number of Lyman limit systems
arising in optically observed galaxies, but this generally involves assuming
a cross section for absorption around a galaxy as a function of its optical
luminosity (Linder et al.~2003; Steidel 1995; Bergeron \& Boiss\'e 1991).  
However the relationship between optical and HI properties of
galaxies may not be well enough understood to make such estimates.
Here we can estimate the number of Lyman limit systems arising from galaxies
that obey an observed HI mass function instead.

We estimate the number of Lyman limit systems arising in each scenario by
putting random lines of sight through the sphere in which the galaxies
are simulated and finding the column density where a line of sight intersects
a galaxy.  
We simulate 20,000 galaxies in an eighth of a sphere having a radius of 108
Mpc, and use 10,000 lines of sight for each case.  The number of Lyman limit systems is then
calculated by correcting to a number density of simulated objects which 
gives rise to
an HI mass function having a normalization consistent with Zwaan et al.~(1997).
The minimum galaxy mass of $10^7 M_\odot$ is used for simulated galaxies, as the
lowest mass objects are not likely to make a substantial contribution to Lyman
limit (or lower column density) absorber counts (Linder 1998), although estimates
from HI studies suggest that slightly more massive objects do contribute to 
Lyman limit absorption (Ryan-Weber et al.~2003).  Values for
the number of Lyman limit absorbers per unit redshift along a line of sight,
$(dN/dz)_{0,LL}$, are shown for the three main simulations in Table~\ref{table1}.

 The number of Lyman limit absorbers has been measured
at redshifts $\ge 0.36$, and the evolution is seen to be about flat or slightly
decreasing down to redshift zero. The lowest redshift values available are within
the range of $(dN/dz)_{LL}\sim 0.2$ to 1.3 (Lanzetta, Wolfe, \& Turnshek
1995b; Storrie-Lombardi et al. 1994; Stengler-Larrea et al. 1995).
Most of the simulations appear to be consistent with these observations, 
although Case
A gives rise to too many absorbers.  The simulation where the ionizing background
intensity was increased gives rise to too few (0.08) absorbers per unit redshift,
but the 
ionizing intensity used is probably more relevant at $z\sim 1$, 
where there are actually more absorbers because less cosmological 
expansion has occurred.

Case A (and some other simulations giving more strongly peaked $N_{HI}^0$ distributions) 
appear to give rise to a somewhat excessive numbers of Lyman 
limit systems.  However, the same problem seems to arise when estimating the 
number of Lyman limit systems around galaxies whose optical luminosity function
is known.  It has long been thought
that luminous galaxies have sufficient cross sections to explain the Lyman 
limit absorber counts fully, yet it is not known why dwarf and LSB
galaxies would not also make some contribution, especially now that such faint
objects are often found to give rise to damped Ly$\alpha$ absorption.
It is possible, for example, that feedback processes change the column density
profiles in the outer parts of some galaxies (McLin, Giroux, \& Stocke 1998).  
While we do not rule
out cases simply because they give rise to somewhat excessive Lyman limit absorber counts,
Case A also has an excessively peaked distribution of inferred column densities, which
is not improved when taking uncertainties in measuring the galaxy radii into consideration.
Thus allowing for a wider range of galaxy column density profiles could make 
Lyman limit absorber counts more consistent with what we expect from observed galaxies.

\section{Conclusions}
Most galaxies have inferred column densities around $10^{20.6\pm 0.3}$ cm$^{-2}$ because
inferred column densities are found by averaging column density profiles, which are
exponential or similar, over a radius where the minimum column density is $\sim 10^{20}$
cm$^{-2}$.  Ionization plays some role in making lower column density objects undetectable, including
those without substantial optical counterparts.  However inferred column density distributions
tell us little about the distribution of central column densities in galaxies and clouds.

Ionization by the background of ultraviolet photons will strongly affect the amount 
of neutral gas remaining, and thus
the HI flux detected, in objects having low hydrogen column densities, if such objects having
sizes comparable to galaxies exist.  Typical HI fluxes are reduced, as a result
of ionization, by a factor of
$\sim 100$ for galaxies having peak column densities $N_{max}\sim 10^{19.5}$ cm$^{-2}$  compared
to those with $N_{max}\sim 10^{20}$ cm$^{-2}$, even if the lower column density galaxies are 
extended in size and just as massive as the higher column density galaxies.

We do not always know the central column densities of the faintest HI sources, but 
the detected inferred column densities are also likely to be above $\sim 10^{20}$ cm$^{-2}$
for most observable galaxies.  Inferred column densities are rather weakly related to 
central column densities for objects having exponential profiles.  Furthermore, since HI 
profiles tend to be mapped out to limiting 
column densities  $\sim 10^{20}$ cm$^{-2}$, it may be difficult to estimate the radii,
and thus the inferred column densities in a consistent manner for objects having lower
$N_{max}$ values.  For example, if the radii are underestimated, which might be more likely
to happen for an extended, diffuse galaxy, the inferred column density could be overestimated.
Other selection effects, such as those against objects having low velocity widths, may also
be important in understanding the observed distribution of inferred HI column densities.

The observed distribution of inferred HI column densities, as seen by Minchin et 
al.~(2003), can easily be simulated assuming possible populations of galaxies having a wide
range of size and central column density distributions, and the simulated distributions
are similar to the HIDEEP distribution for a wide range of model parameters.  
(Thus the 'Frozen Disc' hypothesis of Minchin et al.~2003 seems to be unnecessary in
explaining these observations.)
However, we are thus given little
constraint on the properties of gas rich objects which have so far escaped detection in
the deepest HI surveys.  Given the effects of ionization, we are unable to rule out the
existence of undetected populations of very faint dwarf galaxies or giant gas clouds, as long
as they have low central column densities.  Such objects could make some contribution to 
Ly$\alpha$ absorption, although a more reasonable number of Lyman limit systems
arises if galaxies have a wide, rather than narrow, range of central column 
densities.

The ionizing background radiation is more intense at redshifts around 1 or 2 than at redshift
zero (Haardt \& Madau 1996), and therefore some of the apparently younger galaxies, such as
LSB galaxies, may have been ionized at these redshifts if they have lower central column 
densities (de Blok et al.~1996), thus slowing their evolution.  Ionization may have also 
affected the formation of dwarf galaxies in certain environments at high redshifts (Efstathiou
1992; Tully et al.~2002), as less dense environments are more likely to be optically thin
to ionizing radiation when the dwarf galaxies formed.  Thus dwarf galaxies may have formed
more easily in rich clusters such as Virgo (Sabatini et al.~2003) and Fornax (Kambas
et al.~2000) than in
more diffuse clusters such as Ursa Major (Trentham \& Tully 2002) and other environments
(Roberts et al.~2003).  Understanding the role that ionization plays is thus important
in testing Cold Dark Matter scenarios and other theories related to galaxy formation.

\label{lastpage}

\end{document}